\documentclass[11pt,a4paper]{article}
\usepackage{jheppub}
\usepackage[T1]{fontenc}
\usepackage[english]{babel}
\usepackage{lmodern}
\usepackage{bm}
\usepackage{amsfonts}
\usepackage{subcaption}
\allowdisplaybreaks

\newcommand*{\cf}{cf.\ }
\newcommand*{\ie}{i.e.\ }
\newcommand*{\eg}{e.g.\ }

\title{Energy-energy correlation in hadronic Higgs decays: analytic results and phenomenology at NLO
}

\author[a,b]{Jun Gao,}
\author[c,e]{Vladyslav Shtabovenko,}
\author[d,e]{Tong-Zhi Yang,}

\emailAdd{jung49@sjtu.edu.cn}
\emailAdd{v.shtabovenko@kit.edu}
\emailAdd{tongzhi.yang@physik.uzh.ch}

\affiliation[a]{INPAC, Shanghai Key Laboratory for Particle Physics and Cosmology, School of Physics and Astronomy, Shanghai Jiao Tong University, Shanghai 200240, China}
\affiliation[b]{Center for High Energy Physics, Peking University, Beijing 100871, China}
\affiliation[c]{Institut f\"ur Theoretische Teilchenphysik, Karlsruher Institut f\"ur Technologie (KIT), 76128 Karlsruhe, Germany}
\affiliation[d]{Department of Physics, University of Z\"urich, CH-8057 Z\"urich, Switzerland}
\affiliation[e]{Zhejiang  Institute of Modern Physics, Department of Physics, Zhejiang University, Hangzhou 310027, China}

\abstract{
In this work we complete the investigation of the recently introduced energy-energy correlation (EEC) function in hadronic Higgs decays at next-to-leading order (NLO) in fixed-order perturbation theory in the limit of vanishing light quark masses. The full analytic NLO result for the previously unknown EEC in the $H \to q \bar{q} + X$ channel is given in terms of classical polylogarithms and cross-checked against a numerical calculation. In addition to that, we discuss further corrections to predictions of 
the Higgs EEC event shape variable, including quark mass corrections, effects of parton shower
and hadronization. We also estimate the statistical error on the measurements of the Higgs EEC at future Higgs factories and compare with the current perturbative uncertainty.}

\preprint{P3H-20-060, TTP20-035, ZU-TH 18/20}

\begin{document}
\maketitle
\flushbottom

\section{Introduction}
\label{sec:introduction} 

In the era of the Large Electron Positron (LEP) collider~\cite{Heister:2003aj,Abdallah:2004xe,Achard:2004sv,Abbiendi:2004qz} at CERN and the Stanford Linear Collider (SLC)~\cite{Abe:1994mf} at SLAC, energy-energy correlation function (EEC)~\cite{Basham:1978bw} never enjoyed the same amount of popularity as the six famous event shape variables, which are thrust~\cite{Brandt:1964sa,Farhi:1977sg}, heavy jet mass~\cite{Clavelli:1981yh}, wide and total jet broadening~\cite{Rakow:1981qn, Ellis:1986ig,Catani:1992jc}, $C$ parameter~\cite{Parisi:1978eg, Donoghue:1979vi} and the jet transition variable $Y_{23}$~\cite{Catani:1991hj}.

Nonetheless, we are currently experiencing an unprecedented amount of theoretical work directed towards a better understanding of this observable in the context of perturbative QCD.
One could even go as far as claiming that we are now living in the ``golden age of EEC''.  Analytic results obtained for EEC in $\mathcal{N}=4$ Supersymmetric Yang-Mills (SYM) and QCD evolve hand in hand, with the former making maximal use of exceptional amount of symmetries encoded in the $\mathcal{N}=4$ SYM Lagrangian and the latter relying on more conventional calculational techniques.
The relevance of $\mathcal{N}=4$ SYM calculations for QCD and collider physics is thoroughly explained in \cite{Henn:2020omi}.

A casual bystander might wonder what makes EEC and EEC-like observables so exceptionally well suited for higher-order analytic investigations, including, but not limited to, fixed-order calculations. After all, as of now none of the six famous event shape variables is known analytically at NLO, while in the case of EEC we already have two QCD NLO~\cite{Dixon:2018qgp,Luo:2019nig} and one $\mathcal{N}=4$ SYM NNLO~\cite{Belitsky:2013xxa,Belitsky:2013bja,Belitsky:2013ofa,Henn:2019gkr} fixed-order results. Collinear and back-to-back regions of EEC in $\mathcal{N}=4$ SYM were investigated in \cite{Kologlu:2019mfz,Korchemsky:2019nzm}, while \cite{Moult:2019vou} introduced a formalism for the subleading power resummation of rapidity logarithms. Furthermore, it is worth noting that by making use of the AdS/CFT duality in $\mathcal{N}=4$ SYM  one can also obtain a strong-coupling limit result for the EEC  \cite{Maldacena:1997re,Hofman:2008ar}. In QCD, the collinear limit of EEC can be understood by using the recently available factorization theorem \cite{Dixon:2019uzg}, which also improves the resummation beyond the leading logarithmic (LL) accuracy \cite{Konishi:1978yx,Konishi:1978ax}. The back-to-back limit features an all-order factorization formula \cite{Moult:2018jzp} that makes use of the transverse-momentum dependent (TMD) factorization \cite{Collins:1981uk,Collins:1981va,Kodaira:1981nh}. In this limit,  resummed predictions are currently known at the N$^3$LL$^\prime$ accuracy \cite{deFlorian:2004mp,Tulipant:2017ybb,Moult:2018jzp,Ebert:2020sfi}. 

The answer to the question raised in the above paragraph lies in the very definition of the energy-energy correlator. As we will see below, the Dirac delta that introduces correlations between energies of partons or final state hadrons can be straightforwardly converted to a loop-momentum dependent (albeit nonlinear) propagator and subjected to the standard methods of computing higher-order corrections, such as integration-by-parts (IBP) reduction~\cite{Chetyrkin:1981qh,Tkachov:1981wb} and differential equations~\cite{Kotikov:1991pm,Kotikov:1990kg,Kotikov:1991hm,Bern:1993kr,Remiddi:1997ny,Gehrmann:1999as}. Moreover, these steps can be carried out using off-the-shelf software packages for loop computations: the specifics of our observable (\eg custom IBP equations for loop integrals with nonlinear propagators) can be encoded in the \textsc{Mathematica} scripts used to invoke the existing tools, so that the tools themselves do not require any modifications. The existence of numerical NNLO results~\cite{DelDuca:2016csb,Tulipant:2017ybb} (making use of the CoLoRFulNNLO method \cite{Somogyi:2006da,Somogyi:2006db,Aglietti:2008fe}) as well  
as the availability of public codes (\eg \textsc{Event 2}~\cite{Catani:1996jh,Catani:1996vz}, \textsc{NLOJet++}~\cite{Nagy:2001fj,Nagy:2003tz}, \textsc{Eerad3}~\cite{Ridder:2014wza}) capable of evaluating the EEC numerically greatly facilitate the cross-checks of new analytic results.

It is important to stress that when speaking of ``EEC'' we do not limit ourselves to the original definition of this event shape variable for electron-positron annihilation to partons via the reaction $e^+ e^- \to q \bar{q} + X$. For example, Transverse-Energy-Energy Correlations (TEEC) \cite{Ali:1984yp} have already been studied
in the context of  proton-proton~\cite{Ali:2012rn} and electron-proton~\cite{Ali:2020ksn} collisions.
The back-to-back limit of TEEC can be investigated using recently obtained factorization theorems for hadron-hadron \cite{Gao:2019ojf} and electron-hadron \cite{Li:2020bub} colliders.

Recent considerations of the three-point~\cite{Chen:2019bpb,Chen:2020adz}, four-point~\cite{Chicherin:2020azt}
and multi-point energy correlators~\cite{Chang:2020qpj,Chen:2020vvp} as well as two-point gravitational energy correlators \cite{Gonzo:2020xza} represent further exciting extensions of the original EEC concept and signalize an increased 
interest of the theorist community in such novel event shape variables.

For phenomenological purposes, EEC can be employed as a tool to determine the value of the strong coupling constant (\cf \eg\cite{Kardos:2020igb} for a recent study)
by comparing the available theoretical predictions to the existing electron-positron collider measurements. In~\cite{Luo:2019nig} it was suggested that a new event shape variable, denoted as the Higgs EEC, could provide an intriguing connection between the strong and the Higgs sectors by defining an observable equally accessible to experimentalists analyzing the data from a future Higgs factory and to theorists calculating the corresponding predictions. Furthermore, this observable could be potentially used for the purpose of $\alpha_s$ determinations from hadronic Higgs decays. A high-energy lepton collider,  
be it  CEPC~\cite{CEPCStudyGroup:2018rmc,CEPCStudyGroup:2018ghi}, ILC~\cite{Behnke:2013xla,Baer:2013cma}, FCC-ee~\cite{Gomez-Ceballos:2013zzn} or CLIC~\cite{Aicheler:2012bya,deBlas:2018mhx}, would be capable of copious production of Higgs bosons in the clean environment of $e^+e^-$-annihilations. It is, therefore, not unreasonable to expect that in the future we might witness a high precision measurement of the Higgs EEC using data collected at a leptonic Higgs factory.

The analytic NLO results presented in~\cite{Luo:2019nig} concerned only the $H \to gg + X$ channel calculated in the Higgs Effective Theory (HEFT)~\cite{Wilczek:1977zn,Shifman:1978zn,Inami:1982xt,Kniehl:1995tn} with massless quarks. The goal of this work is to present analytic results also for the channel $H \to q \bar{q} + X$, thus completing the fixed-order investigation of the Higgs EEC at NLO. Being the largest Higgs decay branching ratio, Higgs decaying into bottom quarks has received much attention from the theory community. For example, the partial decay width of $H \to q \bar{q}$ has been calculated to N$^4$LO~\cite{Baikov:2005rw,Davies:2017xsp,Herzog:2017dtz}, and the fully differential decay width for the same process is known to N$^3$LO~\cite{Anastasiou:2011qx,DelDuca:2015zqa,Mondini:2019gid} for massless quarks and to NNLO~\cite{Bernreuther:2018ynm} for massive quarks. Some interesting results obtained very recently are the calculation of Higgs decaying into two bottom quarks and an additional jet at NNLO~\cite{Mondini:2019vub}, the study of the Higgs decay into four bottom quarks at NLO~\cite{Gao:2019ypl} and the investigation of the thrust distribution for Higgs going into a pair of bottom quarks or gluons plus an additional jet at NLO and approximate NNLO~\cite{Gao:2019mlt}. It is also worth mentioning that the NNLL$'$ resummed results are now available for the 2-jettiness distribution describing Higgs decays into $b \bar{b}$ and $gg$~\cite{Alioli:2020fzf}.

For the sake of clarity, in the following we will denote the $H \to gg + X$ and $H \to q \bar{q} + X$ contributions as $Hgg$ EEC and $Hq\bar{q}$ EEC respectively. The Higgs EEC is then understood to contain both channels. The original observable from~\cite{Basham:1978bw} will be referred to as the standard EEC.

Following~\cite{Luo:2019nig}, we define the Higgs EEC as 
\begin{equation}
 \frac{1}{\Gamma_{\textrm{tot}}} \frac{d \Sigma_H  (\chi)}{d \cos\chi} =
\sum_{a,b} \int \, \frac{2 E_a E_b}{m_H^2} \, \delta( \cos\theta_{ab} -
  \cos\chi) \,  d \Gamma_{a+b+X}, \label{eq:eecdef}
\end{equation}
with $\Gamma_{\textrm{tot}}$ being the total decay width for $H \to \textrm{hardons}$, whereas $d \Gamma_{a+b+X}$ describes the differential decay rate of a Higgs decaying into two hadrons plus anything else. Furthermore, we have  $\cos\theta_{ab} = \hat{\bm{p}}_a \cdot \hat{\bm{p}}_b$, where $(E_a, \bm{p}_a)^T$ and $(E_b, \bm{p}_b)^T$ denote the 4-vectors of the hadrons $a$ and $b$ respectively.
Finally, $\chi$ is the angle between two calorimeters measuring the energies of $a$ and $b$, while $m_H$ stands for the Higgs boson mass. By summing over all available final state hadron pairs $(a,b)$ and weighting their contributions to the energy flow by the product of their energies divided by the square of the Higgs mass, we obtain a differential angular distribution normalized to unit area.

To calculate the Higgs EEC in perturbation theory we replace the hadrons by partons and exclude self-correlations, so that the contributions with $a=b$ are removed from the summation in eq.\,\eqref{eq:eecdef}. The interacting part of the relevant Lagrangian reads
\begin{equation}
\mathcal{L}_{\textrm{int}} = - \frac{1}{4} \lambda H \mathrm{Tr} (G^{\mu \nu} G_{\mu \nu}) + \sum_q \frac{y_q}{\sqrt{2}} H \bar{\psi}_q \psi_q,
\end{equation}
where the first term stems from the HEFT with $\lambda$ being the corresponding Wilson coefficient (known up to $\textrm{N}^4\textrm{LO}$~\cite{Baikov:2016tgj}). The second term is the Standard Model Yukawa interaction for quarks, with $y_q$ being the Yukawa coupling for the quark flavor $q$.

 To facilitate the analytic calculation we choose to work in the massless quark limit, while keeping nonvanishing Yukawa couplings. The top quark contributions are thus omitted and we have only 5 active quark flavors. As has already been observed in~\cite{Gao:2019mlt}, the chiral symmetry of massless QCD ensures that in this approximation there is no interference between the $H \to gg + X$  and $H \to q \bar{q} + X$ channels. The respective operators also do not mix under the renormalization so that both pieces can be treated separately.
Since the gluonic channel has already been computed in~\cite{Luo:2019nig}, our sole remaining task is to calculate the contribution from Higgs decaying to a quark-antiquark pair and one or two additional partons. The 3-parton final state corresponds to the LO result, while the 4-parton states are needed for the NLO.

We normalize the $H q \bar{q}$ EEC contribution with respect to the total decay width for $H \to q \bar{q}$ given by
\begin{equation}
\Gamma_{\textrm{tot}} = \frac{y_q^2 (\mu) m_H C_A}{16 \pi} K(\mu), \label{eq:hdecaytot}
\end{equation}
where $C_A$ stands for the number of colors and $K(\mu)$ encodes higher order corrections in $\alpha_s$. The $K$-factor for $H \to b \bar{b}$ in the limit where the bottom mass is set to zero is currently known at $\mathcal{O}(\alpha_s^4)$~\cite{Baikov:2005rw,Davies:2017xsp,Herzog:2017dtz}, and the full scale dependence up to $\mathcal{O}(\alpha_s^3)$ can be found in~\cite{Chetyrkin:1996sr}. This normalization prescription ensures that $H q \bar{q}$ EEC does not depend on $y_q$, while the dependence on $m_H$ enters only through $\log(\mu/m_H)$ and vanishes for the renormalization scale choice $\mu = m_H$.

Our paper is organized as follows. We describe the technical details of our Higgs EEC calculation for the $H \to q \bar{q} + X$ channel in section \ref{sec:calculation} and subsequently present the obtained analytic results  (including the asymptotic behavior in the collinear and back-to-back limits) in section \ref{sec:fullres}. Section \ref{sec:numerics} explores the phenomenological implications of the $H q \bar{q}$ EEC observable. Finally, our conclusions and possible future extensions of this work are summarized in section \ref{sec:summary}.

\section{Technical framework}
\label{sec:calculation} 

Our calculation essentially follows the path that has already been outlined in~\cite{Dixon:2018qgp} and explained in details in~\cite{Luo:2019nig}, so that we keep the following description short.

\begin{figure}
  \centering
  \begin{subfigure}[b]{0.45\textwidth}
      \includegraphics[width=\textwidth]{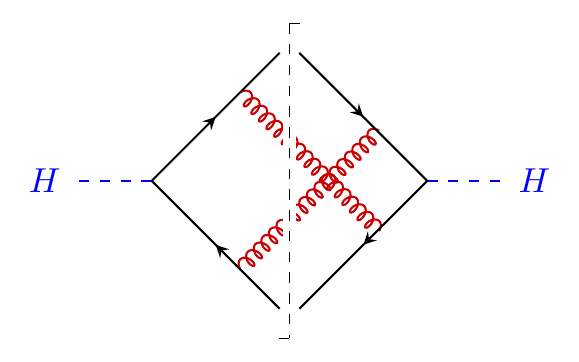}
    \caption{$qqgg$}
    \label{fig:gg}
  \end{subfigure} 
  \begin{subfigure}[b]{0.45\textwidth}
    \includegraphics[width=\textwidth]{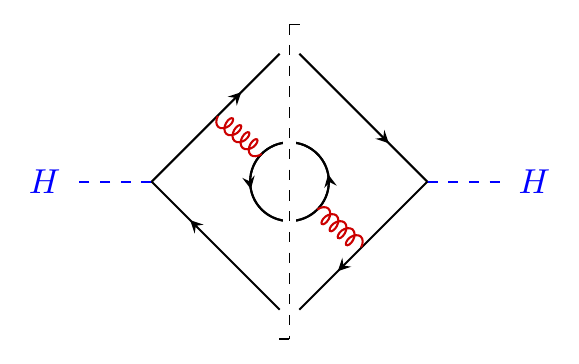}
    \caption{$q\bar{q}q'\bar{q}'$}
    \label{fig:flavor}
  \end{subfigure} 
  \begin{subfigure}[c]{0.45\textwidth}
    \includegraphics[width=\textwidth]{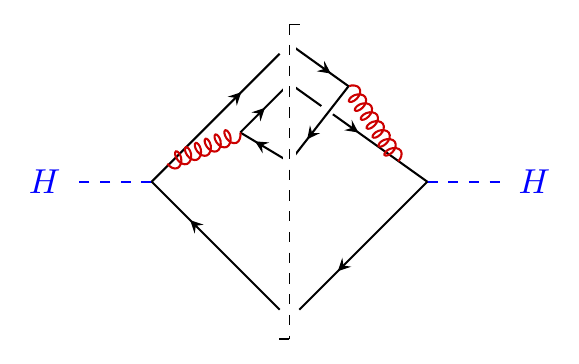}
    \caption{$q\bar{q}q\bar{q}$}
    \label{fig:qq}
  \end{subfigure} 
  \begin{subfigure}[c]{0.45\textwidth}
    \includegraphics[width=\textwidth]{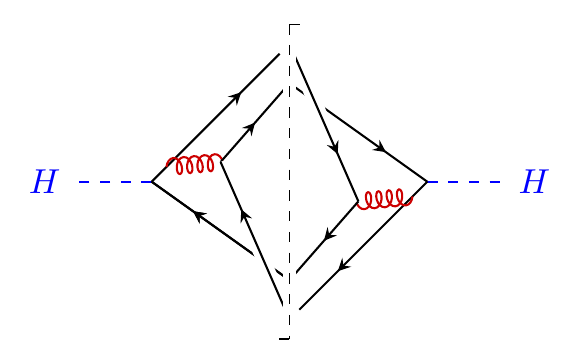}
    \caption{$q\bar{q}q\bar{q}$}
    \label{fig:qq0}
  \end{subfigure}

\caption{Representative cut diagrams for real corrections to the
$H q\bar{q}$ EEC at NLO.}
    \label{fig:realcorr}
\end{figure}

First of all, we need to obtain matrix elements squared $|\mathcal{M}(H \to q \bar{q} + X)|^2$ for real, double-real and real-virtual corrections to the Higgs decaying into a quark-antiquark pair. The real and double-real contributions follow directly from squaring the corresponding tree-level amplitudes with 3- or 4-parton final states respectively
\begin{subequations}
\begin{align}
H (Q) &\to q (p_1) \bar{q} (p_2) g (p_3), \\
H (Q) &\to q (p_1) \bar{q} (p_2) q' (p_3) \bar{q}' (p_4), \\
H (Q) &\to q (p_1) \bar{q} (p_2) q (p_3) \bar{q} (p_4), \\ 
H (Q) &\to q (p_1) \bar{q} (p_2) g (p_3) g (p_4).
\end{align}
\end{subequations}
A visualization of the double-real contributions using the cut diagram notation is shown in figure \ref{fig:realcorr}. Working in the rest frame of the decaying Higgs particle, we have $Q = (m_H,0,0,0)^T$.

The real-virtual piece follows from the interference of the tree-level and 1-loop 3-parton final states. The Higgs EEC observable without the overall normalization factor is obtained by multiplying $|\mathcal{M}(H \to q \bar{q} + X)|^2$ with the measurement function
\begin{equation}
E_a E_b \, \delta (\cos \theta_{ab} - \cos \chi ) =  (p_a \cdot Q)^2 (p_b \cdot Q)^2 \delta \left (
2 z \, p_a \cdot Q \, p_b \cdot Q - p_a \cdot p_b \, Q^2 \right ),
\end{equation}
where we introduced 
\begin{equation}
2 z \equiv 1 - \cos \chi.
\end{equation}

Since the real-virtual piece involves only a massless 3-particle phase space, it is sufficiently simple to be integrated directly via \textsc{HyperInt}~\cite{Panzer:2014caa}. However, the NLO double-real contribution leaves us with a large number of complicated and badly divergent\footnote{The IR safety of the EEC observable guarantees the absence of $1/\varepsilon_{\textrm{IR}}$ poles in the final result but not in the intermediate results.} phase-space integrals. We choose to handle them by employing the method of reverse unitarity~\cite{Anastasiou:2002yz,Anastasiou:2003yy} which effectively trades the measurement function for the following nonlinear cut propagator
\begin{equation}
\frac{1}{2 z \, p_a \cdot Q \, p_b \cdot Q - p_a \cdot p_b \, Q^2} \bigl |_{\textrm{cut}}.
\end{equation}
The occurring loop integrals can then be reduced using IBP techniques. The resulting master integrals can be solved via
differential equations by finding a canonical form~\cite{Henn:2013pwa}
for each of the systems and then determining the integration constants using suitable boundary conditions.

In practice, we generate the Higgs decay amplitudes using \textsc{QGRAF}~\cite{Nogueira:1991ex} and \textsc{FeynArts}~\cite{Hahn:2000kx}. \textsc{FeynCalc}~\cite{Mertig:1990an,Shtabovenko:2016sxi,Shtabovenko:2020gxv}, \textsc{FORM}~\cite{Vermaseren:2000nd} and \textsc{Color}~\cite{vanRitbergen:1998pn} are used to prepare the squared matrix elements, evaluate them in $d$-dimensions and carry out the color algebra. We also employ \textsc{FeynHelpers}~\cite{Shtabovenko:2016whf} and \textsc{Package-X}~\cite{Patel:2015tea,Patel:2016fam} for the calculation of the real-virtual matrix element. To avoid dealing with ghost contributions we make use of the axial gauge
\begin{equation}
\sum_{\lambda=1}^2 \varepsilon^\mu (\bm{p}_i, \lambda) \varepsilon^{\ast \nu} (\bm{p}_i, \lambda) =  -g^{\mu \nu} + \frac{(p_i^\mu n^\nu + p_i^\nu n^\mu)}{p_i \cdot n} - \frac{n^2 p_i^\mu p_i^\nu}{(p_i \cdot n)^2},
\end{equation}
when summing over the gluon polarizations.

The obligatory topology identification step proceeds by considering all possible ways to exchange loop momenta $p_a \leftrightarrow p_b$ or to perform a shift $p_a \to Q - \sum_{b \neq a} p_b$. Notice that the invariance of the sum of the measurement functions for different partons under these manipulations lead to a significant simplification of this task. In the first step, instead of looking at the full integrand 
\begin{equation}
\left (\prod_k \delta_+(p_k^2) \right) |\mathcal{M}(H \to q \bar{q} + X)|^2 \sum_{a<b} \, 2 E_a E_b \, \delta (\cos \theta_{ab} - \cos \chi)
\end{equation}
it is convenient to omit the Dirac delta from the measurement function and enumerate the occurring subtopologies. In the second step we augment each identified subtopology with the corresponding nonlinear cut propagator. In the case of a 4-parton final state, one subtopology gives rise to 6 integral families, stemming from the parton pairs $(1,2)$, $(1,3)$, $(1,4)$, $(2,3)$, $(2,4)$ and $(3,4)$. The subprocesses with $q\bar{q} g$, $q \bar{q} q \bar{q}$ and $q \bar{q} q' \bar{q}'$
final states contain only one subtopology each, given by
\begin{align}
&\{p_1, p_2, Q -p_1 - p_2,Q -p_1, Q -p_2  \}, \\
& \{p_1, p_2, p_3, Q -p_1 - p_2 - p_3, Q -p_1 - p_2, Q -p_1, Q -p_2, p_1 + p_3, p_2 + p_3 \}
\end{align}
and 
\begin{equation}
\{p_1, p_2, p_3, Q -p_1 - p_2 - p_3, Q -p_1, Q - p_3, p_1 + p_2 + p_3, p_1 + p_2, p_1 + p_3 \}
\end{equation}
respectively. The most complicated double-real piece stemming from the $q \bar{q} g g$ final state   involves 3 following subtopologies
\begin{subequations}
\begin{align}
& \{p_1, p_2, p_3, Q -p_1 - p_2 - p_3, Q -p_1 - p_2, Q -p_2 , Q -p_1 , p_1 + p_3, p_2 + p_3\}, \\
&\{p_1, p_2, p_3, Q -p_1 - p_2 - p_3 , Q -p_1 - p_3 , Q -p_2 , Q -p_1, p_1 + p_3, p_1 + p_2 \}, \\
& \{p_1, p_2, p_3, Q -p_1 - p_2 - p_3, Q -p_1 - p_3, Q -p_2 - p_3, Q -p_2, p_1 + p_3,  p_2 + p_3 \}
\end{align}
\end{subequations}
that lead to 18 integral families. The search for a minimal set of subtopologies as well as the generation of the final integral families is done using in-house \textsc{Mathematica} scripts. Custom codes written on top of \textsc{FeynCalc} and \textsc{LiteRed}~\cite{Lee:2012cn} are used to handle linearly dependent propagators via partial fraction decomposition and to derive symbolic equations for the IBP reduction. Then, the IBP-reduction is carried out with \textsc{FIRE}~\cite{Smirnov:2014hma,Smirnov:2019qkx}, where we submit 
our custom IBP equations to the program via the variable  \texttt{startinglist} and mark all cut propagators through the \texttt{RESTRICTIONS} setting.

Finally, we map the obtained master integrals to the set of  integrals that was calculated in~\cite{Dixon:2018qgp}. Just as in the case of the $H gg$ EEC, we find no new master integrals that cannot be expressed as a linear combination of masters from the standard EEC integral basis at NLO. 

Upon adding all contributions together and carrying out the UV-renormalization of the real-virtual contribution, we end up with a manifestly finite result, as expected from the IR-safe property of the EEC event shape variables.

\section{Analytic results at NLO}
\label{sec:fullres}

The main result of this work is the analytic expression for the $H q \bar{q}$ EEC at $\mathcal{O} (\alpha_s^2)$
given by
\begin{align}
&   \frac{1}{\Gamma_{\textrm{tot}}} \frac{d\Sigma_{H q \bar{q}}(\chi)}{d\cos\chi} \nonumber  \\
& =  \frac{1}{K(\mu)}  \left [
 \frac{\alpha_s(\mu)}{2\pi} A_{H q \bar{q}}(z) + \left(\frac{\alpha_s(\mu)}{2 \pi}\right)^2  \left( (\beta_0 + 6 \, C_F)  \log \frac{\mu}{m_H} A_{H q \bar{q}}(z) + B_{H q \bar{q}}(z) \right) \right ],
 \label{eq:eecnlo}
\end{align}
where $\beta_0 = 11/3 C_A - 4/3 N_f T_f$ and $N_f$ stands for the number of quark flavors. The QCD color factors read $C_A = N_c = 3$, $C_F = (N_c^2-1)/(2 N_c) = 4/3$ and $T_f = 1/2$ with $N_c$ being the number of colors. The overall prefactor $1/K(\mu)$ stems from the normalization prescription given in eq.\,\eqref{eq:hdecaytot}, while $A_{H q \bar{q}}(z)$ and $B_{H q \bar{q}}(z)$ denote the LO and NLO coefficients respectively. One may wonder why the coefficient of $\log \frac{\mu}{m_H}$ in the numerator of eq.~\eqref{eq:eecnlo} is proportional to $\beta_0+6 C_F$. The origin of this term can be traced back to the usual strong coupling constant renormalization and the additional Yukawa renormalization~\cite{Gehrmann:2014vha,Gao:2019mlt}, 
\begin{align}
y_q^b =  y_q(\mu) \left( 1- \frac{3 C_F}{2 \epsilon} \frac{\alpha_s}{2\pi} + \mathcal{O}(\alpha_s^2)  \right) \,. 
\end{align}
Notice that $K(\mu)$ to order $\mathcal{O}(\alpha_s)$ is given by 
\begin{align}
\label{eq:KmuDefinition}
K(\mu) = 1+ \frac{\alpha_s}{2 \pi} C_F \left(\frac{17}{2}+ 6 \log \frac{\mu}{m_H} \right) + \mathcal{O}(\alpha_s^2)\,.
\end{align}
Using eq.~\eqref{eq:KmuDefinition}, one could expand eq.~\eqref{eq:eecnlo} to $\mathcal{O}(\alpha_s^2)$, obtaining a result with the coefficient of $\log \frac{\mu}{m_H}$ being exactly proportional to $\beta_0$. 

The LO piece is directly proportional to $C_F$ and can be written as
\begin{align}
A_{H q \bar{q}}(z) & = C_F \left(\frac{-18+15 z}{4 (1-z) z^4}+\frac{\left(-9+12 z-3 z^2-z^3\right) \log (1-z)}{2 (1-z) z^5}\right).
\label{eq:ahzexpl}
\end{align}
The NLO coefficient $B_{H q \bar{q}}(z)$ can be decomposed into
\begin{equation}
B_{H q \bar{q}}(z) = C_F^2 B_{H q \bar{q},\text{lc}}(z) + C_F (C_A - 2 C_F) B_{H q \bar{q},\text{nlc}}(z) + C_F N_f T_f B_{H q \bar{q},N_f}(z),
\label{eq:bhzdecomp}
\end{equation}
where $B_{H q \bar{q},\text{lc}}(z)$, $B_{H q \bar{q},\text{nlc}}(z)$ and $B_{H q \bar{q},N_f}(z)$ stand for the leading color, next-to-leading color and the $N_f$ pieces respectively.
The color structure of the NLO coefficient is identical to the one observed in the standard EEC. This is not surprising, as both observables are quark-initiated quantities.

The analytic structure of the color coefficients precisely follows the pattern known from the standard EEC and the $H gg$ EEC. We again find the same set of building block functions $g_i^{(j)}$, were $j$ denotes the pure transcendental weight
\begin{align}
  g_1^{(1)} &= \log (1-z)\,, \nonumber \\
  g_2^{(1)} &=  \log (z)\,,  \nonumber \\
  g_1^{(2)} &= 2 (\text{Li}_2(z)+\zeta_2)+\log ^2(1-z)\,, 
\nonumber\\
g_2^{(2)} & = \text{Li}_2(1-z)-\text{Li}_2(z)\,, \nonumber \\
g_3^{(2)} &= - 2 \, \text{Li}_2\left(-\sqrt{z}\right)
+ 2 \, \text{Li}_2\left(\sqrt{z}\right)
+ \log\left(\frac{1-\sqrt{z}}{1+\sqrt{z}}\right) \log (z) \,,
\nonumber\\
g_4^{(2)} &= \zeta_2 \,, \nonumber \\
g_1^{(3)}  & = -6
\left[ \text{Li}_3\left(-\frac{z}{1-z}\right)-\zeta_3 \right]
- \log \left(\frac{z}{1-z}\right)
 \left(2 (\text{Li}_2(z)+\zeta_2)+\log^2(1-z)\right)
\,,
\nonumber\\
g_2^{(3)} & =  -12
\left[ \text{Li}_3(z)+\text{Li}_3\left(-\frac{z}{1-z}\right) \right]
+ 6 \, \text{Li}_2(z) \log(1-z) + \log^3(1-z) \,,
\nonumber\\
g_3^{(3)} & = 6 \log(1-z) \, (\text{Li}_2(z)-\zeta_2)
- 12 \, \text{Li}_3(z) + \log^3(1-z) \,, \nonumber\\
g_4^{(3)} &= \text{Li}_3\left(-\frac{z}{1-z}\right)
                - 3 \, \zeta_2 \log(z) + 8 \, \zeta_3 \,,\nonumber\\
g_5^{(3)} &=  
- 8 \left[ \text{Li}_3\left(-\frac{\sqrt{z}}{1-\sqrt{z}}\right)
+ \text{Li}_3\left(\frac{\sqrt{z}}{1+\sqrt{z}}\right) \right]
+ 2 \text{Li}_3\left(-\frac{z}{1-z}\right)
+ 4 \zeta_2 \log (1-z) \nonumber \\ 
& +\log \left(\frac{1-z}{z}\right)
     \log^2\left(\frac{1+\sqrt{z}}{1-\sqrt{z}}\right) \,.
\label{eq:gdef}
\end{align}
The coefficients of these functions (except for some coefficients multiplying $g_3^{(2)}$) are rational polynomials of the form
\begin{equation}
\frac{\sum_{i=1}^7 c_i z^i}{(1-z)^m z^k}, \textrm{ with } c_i \in \mathbb{Z}, \quad  0 \leq m \leq 1, \quad  0 \leq k \leq 5  \ \textrm{ and }  m,k \in \mathbb{N}_0.
\end{equation}
Every color component also contains a term proportional to $1/z^{7/2} g_3^{(2)}$ and those pieces are symmetric under $\sqrt{z} \to - \sqrt{z}$, which appears to be a universal feature of EEC observables at NLO~\cite{Belitsky:2013ofa,Dixon:2018qgp,Luo:2019nig}. On the other hand, it is interesting to observe that the highest power of $z$ in the numerators of the rational polynomials is only 7, at variance with 8 in the case of the $Hgg$ EEC and 9 for the standard EEC at NLO. Furthermore, the largest value of the power $k$ being 5 is true also for the standard EEC, while it can go up to 6 for the $Hgg$ EEC. Presumably, these small differences between the $H q\bar{q}$ EEC and the standard EEC can be largely attributed to the different vertex structures of $\mathcal{L}_{\textrm{int}}$: the former is initiated through a scalar-fermion coupling, while the latter starts via a vector-fermion  interaction.

\begin{figure}[ht]
\centering
\includegraphics[width=0.9\textwidth,clip]{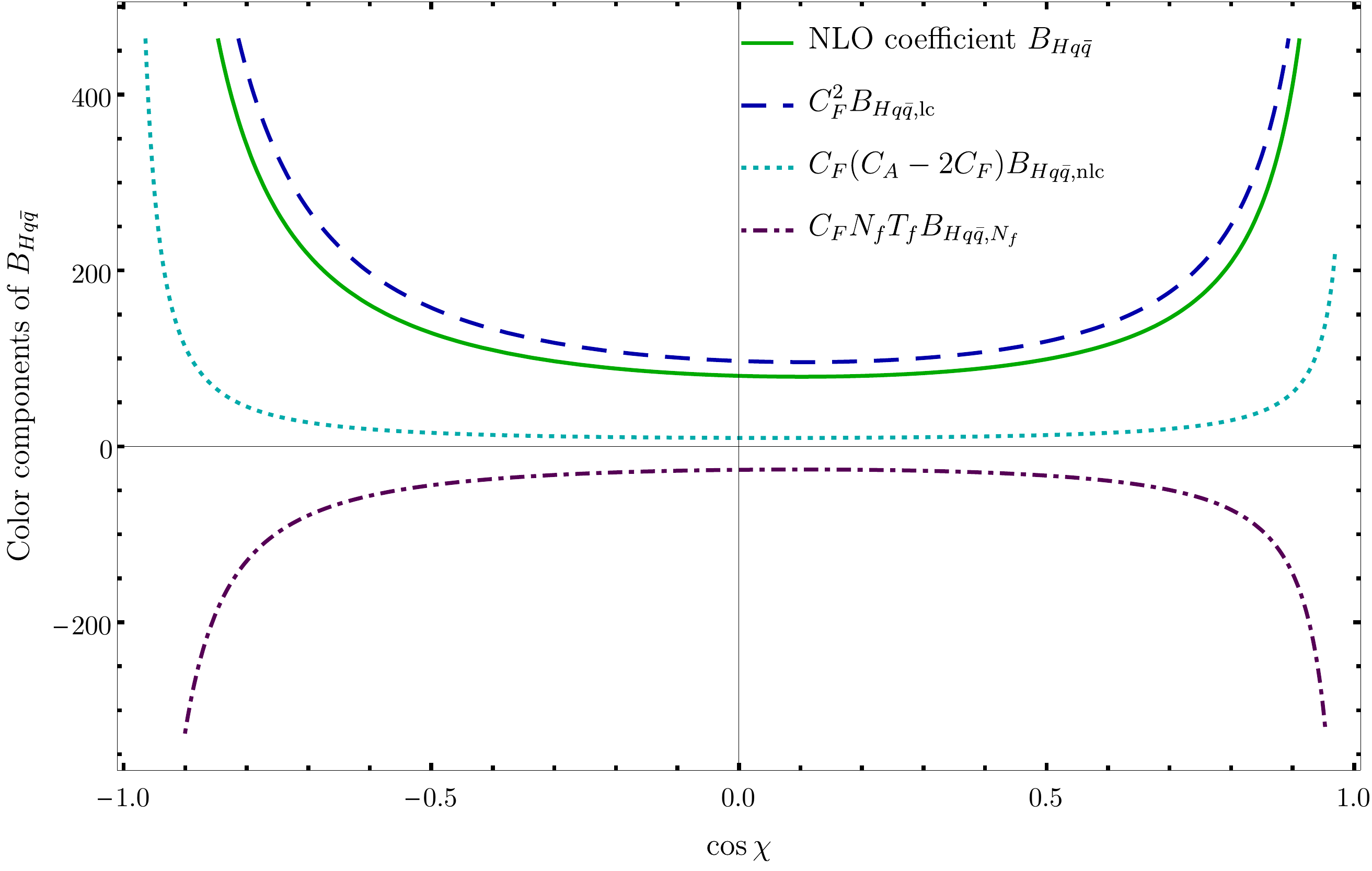}
\caption{NLO coefficient $B_{H q \bar{q}}$ and its color components $B_{H q \bar{q},\textrm{lc}}$, $B_{H q \bar{q},\textrm{nlc}}$ and $B_{H q \bar{q},N_f}$ for $N_f =5$ and $N_c = 3$. Only the $N_f$ piece yields a negative contribution, while both other components contribute positively.}
\label{fig:bhplot}
\end{figure}

The analytic results for the separate color  components at NLO read as follows

\begin{subequations}
\begin{align}
& B_{H q \bar{q},\text{lc}}(z) = -\frac{17422-15003 z-369 z^2-304 z^3+576 z^4-576 z^5}{288 (1-z) z^4} \nonumber \\
&-\frac{\left(4775-9637 z+5189 z^2-387 z^3+436 z^4-1280 z^5+2016 z^6-1152 z^7\right)}{144 (1-z) z^5} g_1^{(1)} \nonumber \\
&+\frac{\left(195+321 z-472 z^2+44 z^3-352 z^4+720 z^5-576 z^6\right) }{72 (1-z) z^4} g_2^{(1)} \nonumber \\
&+\frac{\left(263-195 z-32 z^2+50 z^3+33 z^4-21 z^5\right)}{24 (1-z) z^5} g_1^{(2)} \nonumber\\
& -\frac{\left(65+138 z-94 z^2+32 z^3+64 z^4-96 z^5+192 z^6\right) }{24 z^5}g_2^{(2)}+\frac{(3+35 z) }{96 z^{7/2}} g_3^{(2)}\nonumber \\
& -2 \left(1-2 z+2 z^2\right) g_1^{(3)}  -\frac{\left(19-27 z+10 z^2\right) }{6 (1-z) z^5} g_2^{(3)} +\frac{1}{6 (1-z)} g_3^{(3)}\nonumber \\
&-\frac{\left(461-463 z+168 z^2-26 z^3+48 z^4\right)}{24 (1-z) z^5} g_4^{(2)} ,  \\ \nonumber \\
& B_{H q \bar{q},\text{nlc}}(z) = -\frac{4082-4101 z+471 z^2-137 z^3+288 z^4-288 z^5}{144 (1-z) z^4} \nonumber \\
& -\frac{\left(4610-9529 z+5813 z^2-859 z^3+775 z^4-1604 z^5+2016 z^6-1152 z^7\right) }{144 (1-z) z^5} g_1^{(1)}  \nonumber \\
&-\frac{\left(2496-4245 z+1207 z^2-338 z^3+2056 z^4-2880 z^5+2304 z^6\right) }{288 (1-z) z^4} g_2^{(1)}  \nonumber \\
& +\frac{\left(328-435 z+53 z^2+117 z^3-9 z^4-10 z^5\right) }{48 (1-z) z^5}  g_1^{(2)} \nonumber  \\
&+\frac{\left(208-213 z+36 z^2-11 z^3-118 z^4+96 z^5-192 z^6\right) }{24 z^5} g_2^{(2)} +\frac{\left(291+175 z+384 z^2\right) }{192 z^{7/2}} g_3^{(2)} \nonumber  \\ 
& -\frac{\left(268-428 z+169 z^2+26 z^3-24 z^4\right) }{12 (1-z) z^5} g_4^{(2)} \nonumber  \\
& +\frac{\left(6-33 z+57 z^2-64 z^3+32 z^4\right) }{8 (1-z) z} g_1^{(3)} -\frac{\left(22-39 z+25 z^2-8 z^3+2 z^4-4 z^5\right)}{24 (1-z) z^5} g_2^{(3)} \nonumber  \\
&-\frac{(1-2 z) g_4^{(3)}}{2 (1-z) z} -\frac{\left(3+2 z^2+4 z^3\right) }{8 z^4} g_5^{(3)},  \\ \nonumber \\
& B_{H q \bar{q},N_f}(z) = -\frac{10-277 z+215 z^2+16 z^3}{48 (1-z) z^4}+\frac{\left(381-621 z+321 z^2-53 z^3-24 z^4\right) }{72 (1-z) z^5} g_1^{(1)} \nonumber \\
&+\frac{\left(204-273 z+101 z^2\right) }{48 (1-z) z^4} g_2^{(1)} -\frac{\left(9-12 z+3 z^2+z^3+z^5\right) }{6 (1-z) z^5} g_1^{(2)} -\frac{\left(51-42 z+16 z^2\right) }{12 z^5} g_2^{(2)}  \nonumber \\
&-\frac{(1+5 z) }{32 z^{7/2}}g_3^{(2)} +\frac{\left(87-141 z+70 z^2-12 z^3\right) }{12 (1-z) z^5} g_4^{(2)}.
\end{align}
\end{subequations}
A plot of $B_{H q \bar{q}}(z)$ showing the size of contributions from the three different color components is shown in figure \ref{fig:bhplot}.

The collinear limit of the $H q\bar{q}$ EEC is easily obtained by expanding  the fixed order result around $z=0$. Up to $\mathcal{O}(z)$ this yields
\begin{subequations}
\begin{align}
& A_{H q \bar{q}}(z) =  \frac{1}{z} \frac{3C_F }{8 } +\frac{21 C_F}{40} + \mathcal{O}(z), \\
& B_{H q \bar{q}}(z) = \frac{1}{z} \biggl[\log (z) \left(-\frac{107 C_A C_F}{120}+\frac{53}{240} C_F N_f T_f+\frac{25 C_F^2}{32}\right) \nonumber \\
& +\left(-\frac{25 \zeta _2}{12}+\frac{\zeta _3}{2}+\frac{71677}{10800}\right) C_A C_F-\frac{1217}{900} C_F N_f T_f+\left(\frac{43 \zeta _2}{12}-\zeta _3-\frac{4051}{1728}\right) C_F^2\biggr] \nonumber \\
& +\log (z) \biggl[\left(\frac{21 \zeta _2}{4}-\frac{32089}{3360}\right) C_A C_F+\frac{803 C_F N_f T_f}{2520}+\left(\frac{2029}{180}-\frac{13 \zeta _2}{2}\right) C_F^2\biggr] \nonumber \\
& +\left(\frac{151 \zeta _2}{24}-\frac{65 \zeta _3}{4}+\frac{20108803}{1411200}\right) C_A C_F+\left(-\frac{\zeta _2}{3}-\frac{90047}{66150}\right) C_F N_f T_f \nonumber \\
& +\left(-\frac{33 \zeta _2}{4}+\frac{41 \zeta _3}{2}-\frac{319489}{43200}\right) C_F^2 + \mathcal{O}(z).
\end{align}
\end{subequations}
In the same manner we can also explore the back-to-back limit. Notice that the presence of large logarithms from soft and collinear emissions signals the necessity of a proper resummation using the existing techniques~\cite{Collins:1981uk,Dokshitzer:1999sh,Moult:2018jzp,Gao:2019ojf}. Expanding around $z = 1$ we find
\begin{subequations}
\begin{align}
& A_H(z) = \frac{1}{1-z} \left[-\frac{1}{2} C_F \log (1-z)-\frac{3 C_F}{4}\right]-4 C_F \log (1-z)-\frac{27 C_F}{4} + \mathcal{O}(1-z), \\
& B_H(z) =
\frac{1}{1-z} \biggl[\frac{1}{2} C_F^2 \log ^3(1-z) +\log ^2(1-z) \left(\frac{11 C_A C_F}{12}-\frac{1}{3} C_F N_f T_f+\frac{9 C_F^2}{4}\right) \nonumber \\
& + \log (1-z) \left(\left(\frac{\zeta _2}{2}-\frac{35}{72}\right) C_A C_F+\frac{1}{18} C_F N_f T_f+\left(\zeta _2+\frac{5}{4}\right) C_F^2\right) \nonumber \\
& +\left(\frac{11 \zeta _2}{4}+\frac{3 \zeta _3}{2}-\frac{35}{16}\right) C_A C_F +\left(\frac{3}{4}-\zeta _2\right) C_F N_f T_f+\left(3 \zeta _2-\zeta _3-\frac{27}{16}\right) C_F^2\biggr] \nonumber \\ 
& +\log ^3(1-z) \left[\frac{13 C_A C_F}{24}+\frac{7 C_F^2}{4}\right]  +\log ^2(1-z) \left[\frac{37 C_A C_F}{6}-\frac{4}{3} C_F N_f T_f+\frac{25 C_F^2}{2}\right] \nonumber \\
& +\log (1-z) \left[\left(\frac{41 \zeta _2}{4}-\frac{727}{72}\right) C_A C_F+\frac{103}{36} C_F N_f T_f+\left(\frac{\zeta _2}{2}+\frac{47}{2}\right) C_F^2\right] \nonumber\\
& + \left(\frac{3259 \zeta _2}{96}-\frac{23 \zeta _3}{8}-\frac{27}{2} \zeta _2 \log (2)-\frac{871}{24}\right) C_A C_F +\left(\frac{15 \zeta _2}{16}+\frac{115}{16}\right) C_F N_f T_f \nonumber\\
& + \left(\frac{83 \zeta _2}{24}+\frac{111 \zeta _3}{4}+27 \zeta _2 \log (2)-\frac{2111}{96}\right) C_F^2 + \mathcal{O}(1-z),
\end{align}
\end{subequations}
where the leading power terms can be also obtained using the formalism of~\cite{Moult:2018jzp}.

\section{Phenomenological applications}
\label{sec:numerics}

In the following we present a brief discussion on phenomenological applications of
the $H q\bar{q}$ EEC event shape variable in Higgs boson decays.

\begin{figure}[ht]
\centering
\includegraphics[width=0.9\textwidth,clip]{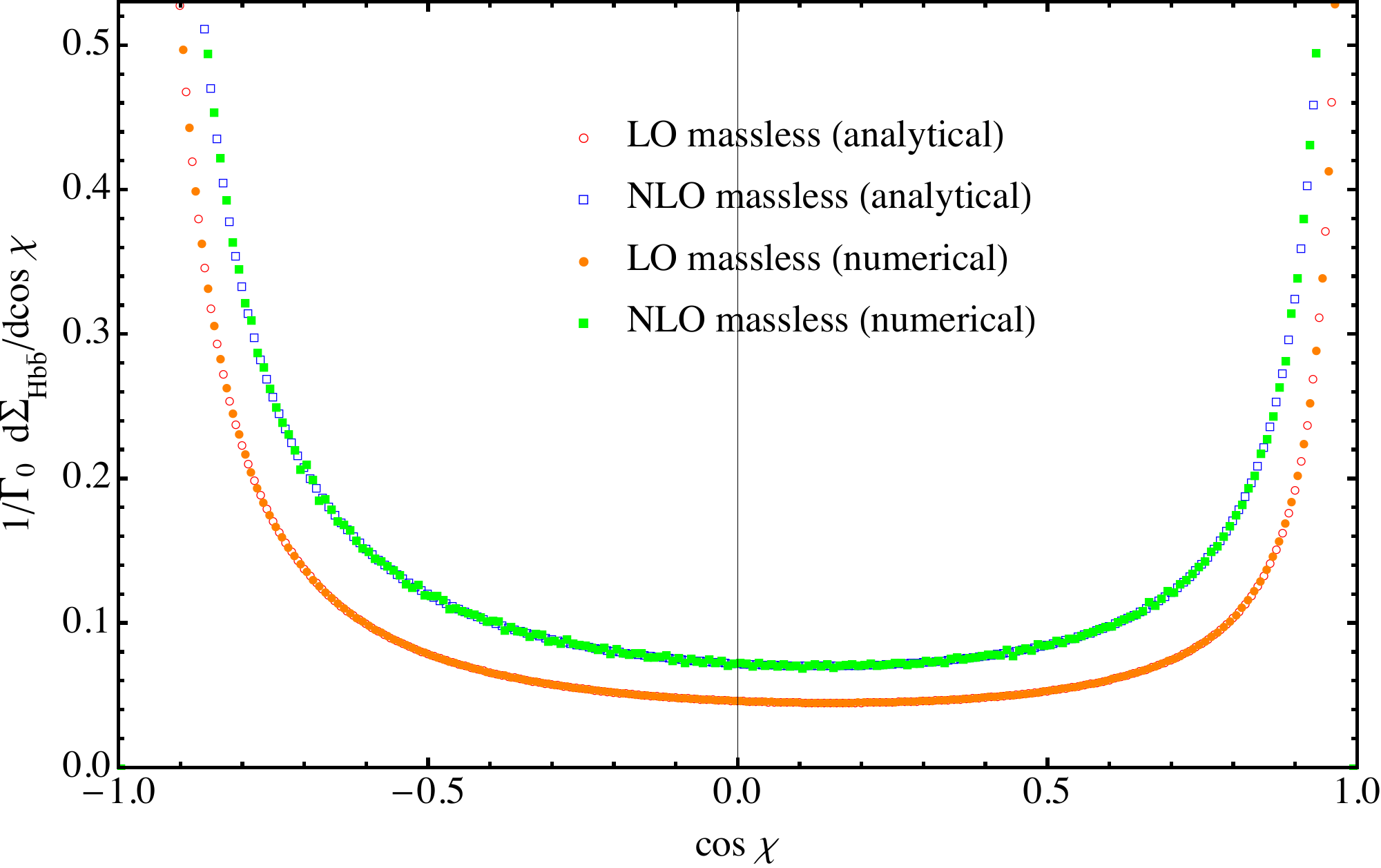}
\caption{
Comparisons between full analytic LO and NLO results for the $H q\bar{q}$ EEC
and the corresponding numerical calculations using MC integrations. 
We consider only Higgs bosons decaying into massless quarks and
normalize each distribution to the total partial width at LO. The MC
errors are much smaller than the size of the markers.
}
\label{fig:plotbench}
\end{figure}

We verify our analytic formulas by comparing them to a numerical result
that was obtained using Monte Carlo (MC) integration. In the numerical calculation we used independent matrix elements that were automatically generated with \textsc{GoSam} 2.0~\cite{Cullen:2014yla}, while the real corrections
were treated using the dipole subtraction method~\cite{Catani:2002hc}. We set the strong coupling constant to
$\alpha_s(M_Z)=0.1181$ in the calculations. Analytic and numerical results for the $H q\bar{q}$ EEC at LO and NLO are shown in figure \ref{fig:plotbench}, where the underlying process is the decay of the Higgs into massless quarks and all distributions are normalized to the total partial width at LO. To simplify the comparison and improve the visual quality of the plot, we choose slightly different $\cos \chi$ values for the curves describing the analytical and numerical distributions. As can be inferred from the plot, within the MC errors we find a perfect agreement between our analytic and numerical predictions both at LO and NLO.

\begin{figure}[ht]
\centering
\includegraphics[width=0.8\textwidth,clip]{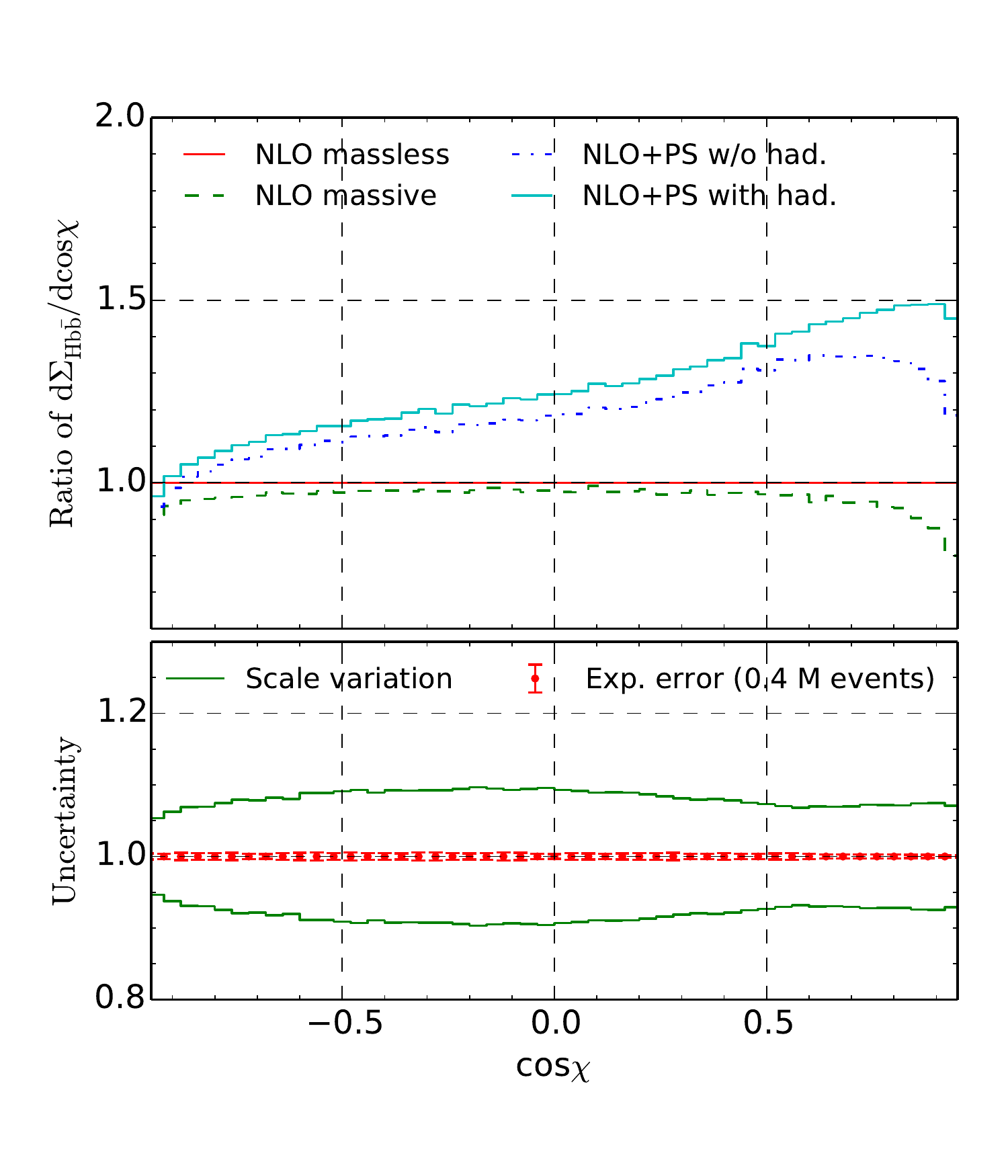}
\caption{
Upper panel: different NLO results for the $H q\bar{q}$ EEC,
where everything is normalized to the 
fixed-order NLO prediction for massless quarks (solid red curve). 
The dashed green curve shows the same calculation done with massive
quarks, while the dot-dashed blue curve also includes matching to
parton shower (but no hadronization). The effects of massive quarks, 
parton shower and hadronization are simultaneously incorporated into the cyan solid curve. 
Lower panel: scale variations of the NLO prediction with massive bottom quarks matched
to parton shower and hadronization and the projected experimental uncertainty.
The latter includes only statistical errors assuming a total number of $4\times 10^5$ events.
}
\label{fig:plotmass}
\end{figure}

A direct comparison to the future experimental data requires additional 
corrections to the fixed-order theory prediction, which we discuss below.

First of all, the effect of the finite bottom-quark mass $m_b$
can be nonnegligible in a fixed-order calculation. Since bottom quarks 
are treated as massless in our analytic result, it is important to estimate 
the impact of this simplification. To this end, we performed another NLO
numerical calculation of the $H q\bar{q}$ EEC, where the bottom-quark
was treated as massive with $m_b = 4.78\textrm{ GeV}$. The ratio of the $H q\bar{q}$ EEC results with massive and massless quarks is shown in the upper panel of figure \ref{fig:plotmass}. We observe that in the range of $|\cos\chi|<0.95$ the bottom-quark mass corrections reduce the distribution by about 4\% in the bulk region and can reach 10\% to 20\% in the back-to-back and collinear regions. The latter is not surprising, as it is well known that collinear radiations are suppressed due to the finite quark mass.

Second, for a meaningful Higgs EEC prediction we must also include parton shower and hadronization corrections. To account for that, we  match our NLO calculation with massive bottom quarks to parton shower using \textsc{POWHEG-BOX-V2}~\cite{Frixione:2007vw,Alioli:2010xd} and \textsc{PYTHIA} 8.2~\cite{Sjostrand:2014zea}. In the \textsc{PYTHIA} setup we use the Monash tune~\cite{Skands:2014pea} and, for the sake of simplicity, force all $B$ hadrons to be stable. Figure \ref{fig:plotmass} shows that the parton shower can substantially enhance the EEC distribution in the whole $\cos\chi$ range, with the corrections amounting to almost 40\% in the collinear region. This observation hints that fixed-order NNLO QCD corrections to the $H q\bar{q}$ EEC could be potentially large. Furthermore, the hadronization 
corrections are equally significant and can reach more than 10\%.

Finally, we estimate the perturbative uncertainty of the matched NLO predictions
by varying the renormalization scale and the square of the parton shower scale independently
by a factor of two around their nominal values, chosen as $m_H$ for the renormalization scale
and $k^2_T$ for the square of the shower scale. We add the two scale variations in quadrature 
and plot the uncertainty band in the lower panel of figure \ref{fig:plotmass}. The total
uncertainty from the scale variations lies between 5\% and 10\% in the plotted region.
Given the existence of NNLO numerical calculations for massless bottom 
quarks~\cite{Mondini:2019vub}, we expect that this uncertainty can be,
in principle, significantly reduced in the future.

In addition to that, the lower panel of figure \ref{fig:plotmass}
also contains the projected experimental uncertainties. In this case we incorporate only the statistical errors and assume the total number of events being $4\times 10^5$, which corresponds to the number of $H \to b \bar{b}$ decays that CEPC~\cite{An:2018dwb} is expected to collect during its first 7 year data taking period. We estimate the statistical errors by first generating 40 ensembles of events
and then calculating the standard deviation of the EEC in each bin from the values
predicted by all ensembles. This procedure is meant to account for the strong statistical correlations
among different bins that typically arise when studying EEC-like observables: 
a single event generates multiple histogram entries, hence simultaneously contributing 
to many bins. In our case, for all bins the observed uncertainties constitute at most 0.5\%.

Experimental systematic errors, that we choose to ignore here, can be
attributed to the signal extraction from the SM background as
well as event reconstruction and detector resolution.
Although they are expected to be dominant over the statistical errors,
a thorough estimation of these uncertainties is beyond the scope of the present paper.

\section{Summary}
\label{sec:summary}
Higgs EEC is a novel event shape variable that can be measured by reconstructing 4-vectors of the final state particles originating from hadronic Higgs decays. This observable opens an interesting perspective of $\alpha_s$ determinations from Higgs precision measurements at future Higgs factories and is therefore of great relevance for experimentalists interested in exploring Higgs phenomenology at high-energy lepton colliders.

In this work we employed methods pioneered in~\cite{Dixon:2018qgp} to calculate Higgs EEC in the $H \to q \bar{q} + X$ channel at NLO in the fixed-order perturbation theory. This result can be combined with the already available computation in the $H \to g g + X$ channel~\cite{Luo:2019nig} to obtain the full Higgs EEC in the limit of vanishing light quark masses. The analytical structure of the $H q \bar{q}$ EEC is very similar to that of the $H g g$ EEC and the standard EEC: all 3 results can be calculated using the same set of master integrals and written in terms of the same building block functions that involve classical polylogarithms up to weight 3.

As far as the phenomenology of the $H q \bar{q}$ EEC is concerned, 
we employed numerical methods to study the importance of the effects
missing in the analytic calculation: finite bottom-quark mass, parton shower and hadronization.
On the one hand, the corrections due to the finite bottom-quark mass turn out to be 
numerically rather small, apart from the collinear and back-to-back regions.
On the other hand, parton shower and hadronization effects can lead to enhancements of tens of percents
beyond the NLO fixed-order predictions. The remaining scale variations are at the level of 5\% to 10\%.
At the same time, the projected statistical uncertainties on the measurements of the Higgs EEC at future Higgs factories
are at sub-percent level. Therefore, we conclude that improved perturbative calculations and a more accurate modeling of the hadronization are mandatory in order to match the future experimental precision.

Theoretical investigations of EEC-like observables continue to expand our understanding of the mathematical underpinning of perturbative QCD. 
The multitude of results made available in the recent years corroborate that the study of EEC has become a very active field of research within the phenomenology of the strong interactions at high energies. Even though every new calculation raises the bar a bit higher, there is obviously still a lot of work left to be done. At NLO one could consider other underlying processes that lead to hadronic decays or try to incorporate effects of massive quarks, while at NNLO we still lack the full fixed-order result even for the standard EEC. Taking a broader view, it would be very rewarding to search for techniques that could enable us to obtain NLO analytic results for event shape variables other than the EEC.
Given the amount of progress in the field made in the last few years, we may very well expect to witness even more exciting findings in the years to come.

\acknowledgments

We are grateful to Ming-xing Luo and Hua Xing Zhu for collaboration at the early stage of this work and important comments on the manuscript. We thank Han-tian Zhang for useful discussions. The work of J.\,G. was sponsored by the National Natural Science Foundation of China under the Grant No. 11875189 and No.11835005. The work of V.\,S. and T.\,Z.\,Y. was supported in part by the National Science Foundation of China (11135006, 11275168, 11422544, 11375151, 11535002) and the Zhejiang University Fundamental Research Funds for the Central Universities (2017QNA3007). V.\,S. also acknowledges the support from the DFG under grant 396021762 -- TRR 257 ``Particle Physics Phenomenology after the Higgs Discovery. T.Z.Y. also wants to acknowledge the support from the Swiss National Science Foundation (SNF) under contract 200020-175595.

\appendix

\section{Asymptotics of the NLO color components}

\subsection{Collinear limit}

\begin{subequations}
\begin{align}
B_{H q\bar{q},\textrm{lc}}(z) &= \frac{13 \zeta _2}{3}-12 \zeta _3+ \frac{1}{z} \left[-\frac{481}{480} \log (z)-\frac{7 \zeta _2}{12}+\frac{472141}{43200}\right]+\left [4 \zeta _2-\frac{7891}{1008}\right] \log (z) \nonumber \\
& +\frac{5583931}{264600} + \mathcal{O}(z), \\ \nonumber \\
B_{H q\bar{q},\textrm{nlc}}(z) &= \frac{151 \zeta _2}{24}-\frac{65 \zeta _3}{4}+ \frac{1}{z} \left[-\frac{107}{120}  \log (z) -\frac{25 \zeta _2}{12}+\frac{\zeta _3}{2}+\frac{71677}{10800}\right] \nonumber \\
& +\left[\frac{21 \zeta _2}{4}-\frac{32089}{3360}\right] \log (z)+\frac{20108803}{1411200} + \mathcal{O}(z), \\ \nonumber \\ 
B_{H q\bar{q},N_f}(z) &=  \frac{1}{z} \left[\frac{53 }{240} \log (z)-\frac{1217}{900}\right]+\frac{803}{2520}  \log (z) -\frac{\zeta _2}{3} -\frac{90047}{66150} + \mathcal{O}(z).
\end{align}
\end{subequations}

\subsection{Back-to-back limit}

\begin{subequations}
\begin{align}
B_{H q\bar{q},\textrm{lc}}(z) &=  \frac{1}{1-z} \biggl [\frac{1}{2} \log ^3(1-z)  +\frac{49}{12} \log ^2(1-z) + \left(2 \zeta _2+\frac{5}{18}\right) \log (1-z)  +2 \zeta _3 \nonumber \\
&  + \frac{17 \zeta _2}{2} -\frac{97}{16} \biggr ]+\frac{17}{6} \log ^3(1-z) +\frac{149}{6} \log ^2(1-z)  +\left[21 \zeta _2+\frac{119}{36}\right] \log (1-z)  \nonumber \\
&  +22 \zeta _3 + \frac{3425 \zeta _2}{48}-\frac{9079}{96} + \mathcal{O}(1-z), \\ \nonumber \\
B_{H q\bar{q},\textrm{nlc}}(z) &=\frac{1}{1-z} \biggl [\frac{11}{12} \log ^2(1-z) +\left( \frac{\zeta _2}{2}-\frac{35}{72}\right) \log (1-z) +\frac{3 \zeta _3}{2} + \frac{11 \zeta _2}{4} -\frac{35}{16} \biggr ] \nonumber \\ 
& +\frac{13}{24} \log ^3(1-z) +\frac{37}{6} \log ^2(1-z) +\left[\frac{41 \zeta _2}{4}-\frac{727}{72}\right] \log (1-z) \nonumber  \\
& -\frac{23 \zeta _3}{8}-\frac{27}{2} \zeta _2 \log (2) + \frac{3259 \zeta _2}{96} -\frac{871}{24} + \mathcal{O}(1-z), \\ \nonumber \\ 
B_{H q\bar{q},N_f}(z) &=  \frac{1}{1-z} \biggl [ -\frac{1}{3} \log ^2(1-z)+\frac{1}{18} \log (1-z) -\zeta _2 +\frac{3}{4} \biggr ] -\frac{4}{3} \log ^2(1-z) \nonumber \\
& +\frac{103}{36} \log (1-z) + \frac{15 \zeta _2}{16} +\frac{115}{16} + \mathcal{O}(1-z).
\end{align}
\end{subequations}

\section{Identical-quark interference contributions}

In this section we provide results for the identical-quark interference contribution to the $H q\bar{q}$ EEC at NLO. Such results are already available for the standard EEC~\cite{Dixon:2018qgp} and the $Hgg$ EEC~\cite{Luo:2019nig}, so that it is useful to have them also for the  $H q\bar{q}$ EEC. The interference terms correspond to the $q\bar{q} q \bar{q}$ cut diagram from figure \ref{fig:qq}, which contributes to $B_{H q \bar{q}\,,\text{nlc}}$. In addition to that, $B_{H q \bar{q}\,,\text{nlc}}$ also receives contributions from the $q \bar{q} gg $ cut diagram from figure \ref{fig:gg} (denoted as $B_{H q \bar{q},\,g}$) and the real-virtual diagrams $B_{H q\bar{q},\text{V}}$. We can, therefore, decompose $B_{H q \bar{q}\,,\text{nlc}}$ as

\begin{align}
B_{H q \bar{q},\,\text{nlc}}(z) = B_{H q \bar{q},\,g} + B_{H q\bar{q},\text{V}}  + B_{H q \bar{q},\,q q_{\text{int}}}    \,.
\end{align}
Since $B_{H q \bar{q}\,,q q_{\text{int}}}$  receives no virtual corrections at NLO, this piece is separately gauge invariant and IR finite. It can be written as
\begin{align}
\label{eq:QQint}
B_{H q \bar{q},q q_{\text{int}}}  &= \frac{-1726+991 z+347 z^2+136 z^3-558 z^4+576 z^5}{288 (1-z) z^4} \nonumber 
\\ 
&+\frac{\left(373-938 z+548 z^2+28 z^3-785 z^4+828 z^5-1152 z^6\right) }{144 z^5} g_1^{(1)} \nonumber 
\\
&-\frac{\left(-1398+1161 z-589 z^2-111 z^3+1055 z^4-1404 z^5+1152 z^6\right)}{144 (1-z) z^4}  g_2^{(1)}  \nonumber 
\\
&+\frac{\left(-8+26 z-47 z^2+24 z^3+23 z^4\right) }{48 z^5} g_1^{(2)} +\frac{\left(12+16 z+35 z^2+48 z^3\right) }{24 z^{9/2}} g_3^{(2)} \nonumber 
\\
&-\frac{\left(-19+126 z+7 z^2+28 z^3+127 z^4-90 z^5+192 z^6\right) }{24 z^5} g_2^{(2)} \nonumber 
\\
&-\frac{\left(11-99 z+52 z^2+66 z^3-48 z^4\right) }{24 (1-z) z^5} g_4^{(2)} +\frac{2 \left(30-25 z+7 z^2\right) }{3 \sqrt{1-z} z^{9/2}} g_5^{(2)} \nonumber 
\\
&
+\frac{1}{8} \left(-26+\frac{6}{z}+31 z-32 z^2\right) g_1^{(3)}
-\frac{(1-2 z) }{2 (1-z) z} \left(\frac{1}{6} g_2^{(3)}+g_4^{(3)}+\frac{1}{2} g_5^{(3)}\right)  \nonumber 
\\
&+\frac{\left(19-11 z+2 z^2\right)}{8 z^5}  \left(-\frac{1}{2} g_2^{(3)}+g_5^{(3)}-\frac{1}{2} g_6^{(3)}+g_7^{(3)}\right)  \nonumber 
\\
&-\frac{\left(-10+15 z-7 z^2+z^3\right)}{12 (1-z) z^5} \left(-3 g_5^{(3)}-33 g_8^{(3)}+g_9^{(3)}\right),
\end{align}
where we introduced some additional building block functions beyond those given in eq.\,\eqref{eq:gdef}. These functions are defined as follows
\begin{align}
 \label{eq:evenMoreGfunc}
  g_{5}^{(2)} = \; & \frac{1}{i} \left[ \text{Li}_2(i r) - \text{Li}_2(-i r)
        - \log(r) \log\left(\frac{1+i r}{1-i r}\right) \right] \,,
\nonumber \\
  g_{6}^{(3)} = \; & \log ^3(1-z)-15 \zeta_2 \log (1-z) \,,
\nonumber \\
  g_{7}^{(3)} = \; & \log (1-z) \left(\text{Li}_2(z)+\log (1-z) \log (z)-\frac{15 \zeta_2}{2}\right) \,,
\nonumber \\
  g_{8}^{(3)} = \; & \zeta_3 \,,
\nonumber \\
  g_{9}^{(3)} = \; & -12 \bigg[
 -\text{Li}_3\left(\frac{1}{2} (1-i r)\right)
 -\text{Li}_3\left(\frac{1}{2} (1+i r)\right)
 +\text{Li}_3(-i r)+\text{Li}_3(i r)
 +\text{Li}_3\left(-\frac{2 r}{i-r}\right)
 \nonumber 
 \\
  &+\text{Li}_3\left(\frac{2 r}{i+r}\right) - \zeta_3 \bigg]
+ 3 \, \text{Li}_3\left(-\frac{z}{1-z}\right)
 + 2 \left[ \log^3\left(\frac{1}{2} (1-i r)\right)
          +\log^3\left(\frac{1}{2} (1+i r)\right) \right]\nonumber \\
&
 - 3 \, (2 \log (i r)-i \pi ) \, \log^2 \left(\frac{1-i r}{1+i r}\right)
-\pi ^2 \left(\log \left(\frac{1}{2} (1-i r)\right)+\log \left(\frac{1}{2} (1+i r)\right)\right) \,,
\end{align}
where $r = \sqrt{z}/\sqrt{1-z}\,.$ The function $g^{(2)}_5$ is real-valued in $z = i r, \bar{z} = -i r$ and is known as the Bloch-Wigner function. The building block functions in eq.~\eqref{eq:evenMoreGfunc} and their specific combinations in eq.~\eqref{eq:QQint} are identical to those appearing in the  $B_{q q_{\text{int}}}$ term of the standard EEC~\cite{Dixon:2018qgp}. This is not surprising, since both the standard EEC and the $H q \bar{q}$ EEC are quark-initiated observables. 

As far as the asymptotics is concerned, in the collinear limit $B_{H q \bar{q}\,,q q_{\text{int}}}$ reads 
\begin{align}
\label{eq:qqintzto0}
 B_{H q \bar{q}\,,q q_{\text{int}}} &=
\frac{1}{z}\left( \frac{\zeta
   _3}{2} -\frac{43 \zeta _2}{24}+\frac{8011}{3456}\right) +\left(5 \zeta
   _2-\frac{59011}{7200}\right) \log (z) \nonumber \\
 &  -\frac{31 \zeta _3}{2} +\frac{727 \zeta
   _2}{120}+\frac{3711491}{432000}+ \mathcal{O}(z). 
\end{align}
We observe that the leading power terms (\ie  the coefficient of $1/z$ in eq.~\eqref{eq:qqintzto0}) are the same as the corresponding terms of the standard EEC~\cite{Dixon:2018qgp}. These terms can be predicted by the jet calculus approach at the next-to-leading logarithm (NLL) accuracy~\cite{Campbell:1997hg}. 

Looking at the back-to-back limit by expanding $B_{H q \bar{q}\,,\text{qqint}}$ up to next-to-leading power
\begin{align}
B_{H q \bar{q}\,,q q_{\text{int}}} &=\frac{1}{1-z} \left( -\frac{\zeta
   _3}{2}+ \frac{3 \zeta _2}{4}-\frac{13}{16} \right)+\left(\frac{51 \zeta
   _2}{4}-\frac{137}{8}\right) \log (1-z)+\frac{5}{8} \log
   ^2(1-z) \nonumber 
   \\
   & -34 \zeta _3-15 \zeta _2
   \log (2)+\frac{889 \zeta _2}{24}+\frac{557}{96} + \mathcal{O}(1-z) \,,
\end{align}
we again find that the coefficient of $1/(1-z)$ precisely reproduces the corresponding results for the standard EEC~\cite{Dixon:2018qgp}.

\bibliographystyle{jhep}
\bibliography{inspire.bib}

\end{document}